\documentclass[prl,twocolumn,showpacs]{revtex4}

\usepackage{graphicx}
\usepackage{dcolumn}
\usepackage{bm}

\begin{document}

\title{Noise activated switching in a driven, nonlinear micromechanical oscillator}

\author{C. Stambaugh}
\author{H. B. Chan}
\email{hochan@phys.ufl.edu}
\affiliation{Department of Physics, University of Florida, Gainesville, FL 32611}

%
\begin{abstract}
We study noise induced switching in systems far from equilibrium by using an underdamped micromechanical torsional oscillator driven into the nonlinear regime.  Within a certain range of driving frequencies, the oscillator possesses two stable dynamical states with different oscillation amplitudes. We induce the oscillator to escape from one dynamical state into the other by introducing noise in the excitation. By measuring the rate of random transitions as a function of noise intensity, we deduce the activation energy as a function of frequency detuning. Close to the critical point, the activation energy is expected to display system-independent scaling. The measured critical exponent is in good agreement with variational calculations and asymptotic scaling theory.  
\end{abstract}

\maketitle


Fluctuation-induced escape from a metastable state is an important problem that is relevant to many phenomena, such as protein folding and nucleation in phase transitions. For systems in thermal equilibrium, the escape rate can be deduced from the height of the free-energy barrier \cite{1}. The barrier decreases as the control parameter $\mu$  approaches a critical (bifurcational) value $\mu_c$ where the metastable state disappears. It has been established theoretically and experimentally \cite{2,3} that, in the simplest and arguably most common case of the saddle-node (spinodal) bifurcation \cite{4}, the barrier height scales as $(\mu-\mu_c)^{3/2}$. Much less is known about escape in systems far from thermal equilibrium \cite{5,6,7,8}. Such systems are not characterized by free energy, and the scaling behavior of the escape rate near a saddle-node bifurcation has not been studied experimentally. Recently the problem of escape far from equilibrium has attracted significant experimental attention in the context of systems where multistability itself arises as a result of strong periodic modulation. Escape was studied in parametrically driven electrons in a Penning trap \cite{9}, doubly clamped nanomechanical oscillators \cite{10,11} and radio frequency driven Josephson junctions \cite{12}.

\begin{figure}[ht]
\includegraphics[scale=.41,angle=0]{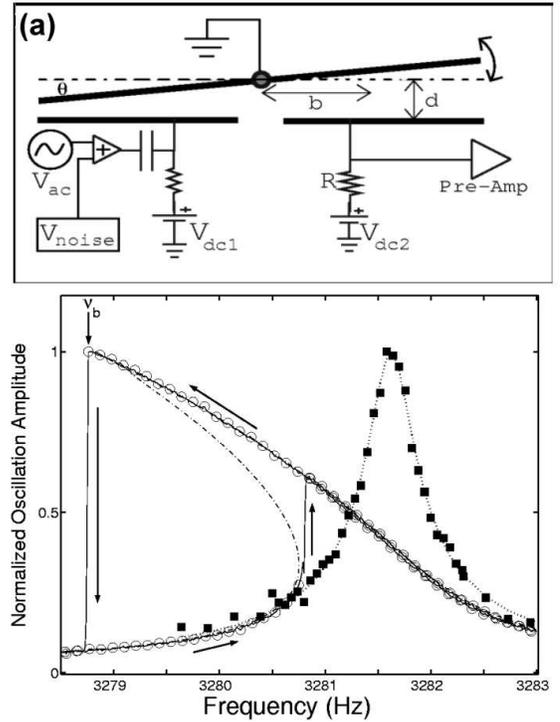} 
\caption{\label{fig:1}    (a) A cross-sectional schematic of the micromechanical torsional oscillator with electrical connections and measurement circuitry (not to scale). (b) Normalized frequency response of sample A for excitation voltages amplitudes of $9.1\;\mu V$ (squares) and $86.5\;\mu V$ (circles). The dotted line represents a fit to the data at smaller excitation using the response of a damped harmonic oscillator. For the large excitation, two dynamical states coexist from $3278.76\;Hz$ to $3280.8\;Hz$. The dashed line fits the data to a damped oscillator with cubic non-linearity \cite{17}, yielding $\beta=-1.2\times10^{14}\; rad^{-2}\: s^{-2}$.
}
\end{figure}  

We report here our investigation of noise-activated switching in systems far from equilibrium.  By using a well-characterized system, an underdamped micromechanical torsional oscillator periodically driven into nonlinear oscillations, we study the dependence of the escape rate on the control parameter as it approaches the critical value and reveal the scaling of the activation energy of escape in a system far from thermal equilibrium. The strongly-driven micromechanical oscillator has two stable dynamical states with different oscillation amplitude within a certain range of driving frequencies. We induce the oscillator to escape from one state into the other by injecting noise in the driving force. By measuring the rate of random transitions as a function of noise intensity, we demonstrate the activated behavior for switching and deduce the activation energy as a function of frequency detuning. Close to the bifurcation frequency where the high-amplitude state disappears, the activation energy is predicted by variational calculations and asymptotic scaling theory to display system-independent scaling \cite{13,14,16}. Our measurement of activation energy as a function of frequency detuning yields critical exponents of $1.38\pm0.15$  and $1.40\pm0.15$  respectively for two samples with different parameters, in good agreement with the theoretical predictions for escape in non-equilibrium systems near a spinodal point.

In our experiment, measurements were performed on two micromechanical oscillators (samples A and B). Each oscillator consists of a moveable polysilicon plate ($500 \; \mu m$ by $500 \; \mu m$ by $3.5\; \mu m$) that is supported by two torsional rods ($40 \; \mu m$ by $4 \; \mu m$ by $2\; \mu m$). The other ends of the torsional rods are anchored to the silicon substrate. A $2$-$\mu m$-thick sacrificial silicon oxide layer beneath the plate was etched away, leaving the plate suspended between the torsional rods. The moment of inertia $I$ for the plate is  $4.3\times10^{-17}\; kg \: m^2$, while the torsional spring constants are $1.83\times10^{-6}\; N \: m \: rad^{-1}$ and  $1.85\times10^{-6}\; N\: m \: rad^{-1}$ for samples A and B respectively. Beneath the top plate, there are two fixed electrodes ($500 \;\mu m$ by $250\; \mu m$) on each side of the torsional springs. One of the electrodes is used for exciting the torsional oscillations while the other electrode is used for detecting the oscillations. More details about the oscillators can be found in Refs. \cite{18, 19}.

Figure \ref{fig:1}a shows a cross-sectional schematic of the oscillator with electrical connections and measurement circuitry. The application of a periodic voltage with DC bias $V_{dc1}$ to one of the electrodes leads to an electrostatic attraction between the grounded top plate and the electrode. Torsional oscillations of the top plate are excited by the periodic component of the electrostatic torque. The detection electrode is connected to a dc voltage $V_{dc2}$ through a resistor $R$. As the plate oscillates, the capacitance between the plate and the detection electrode changes. The detection electrode is connected to a charge sensitive preamplifier followed by a lock-in amplifier that measures the signal at the excitation frequency. Measurements for both samples were performed at pressure of less than $2\times10^{-7}\; torr$. Sample A was measured at liquid nitrogen temperature and sample B was measured at liquid helium temperature. The quality factors $Q$ are about $2,000$ and $8,000$ for samples A and B respectively at the corresponding measurement temperatures.

\begin{figure}[ht]
\includegraphics[scale=.21,angle=0]{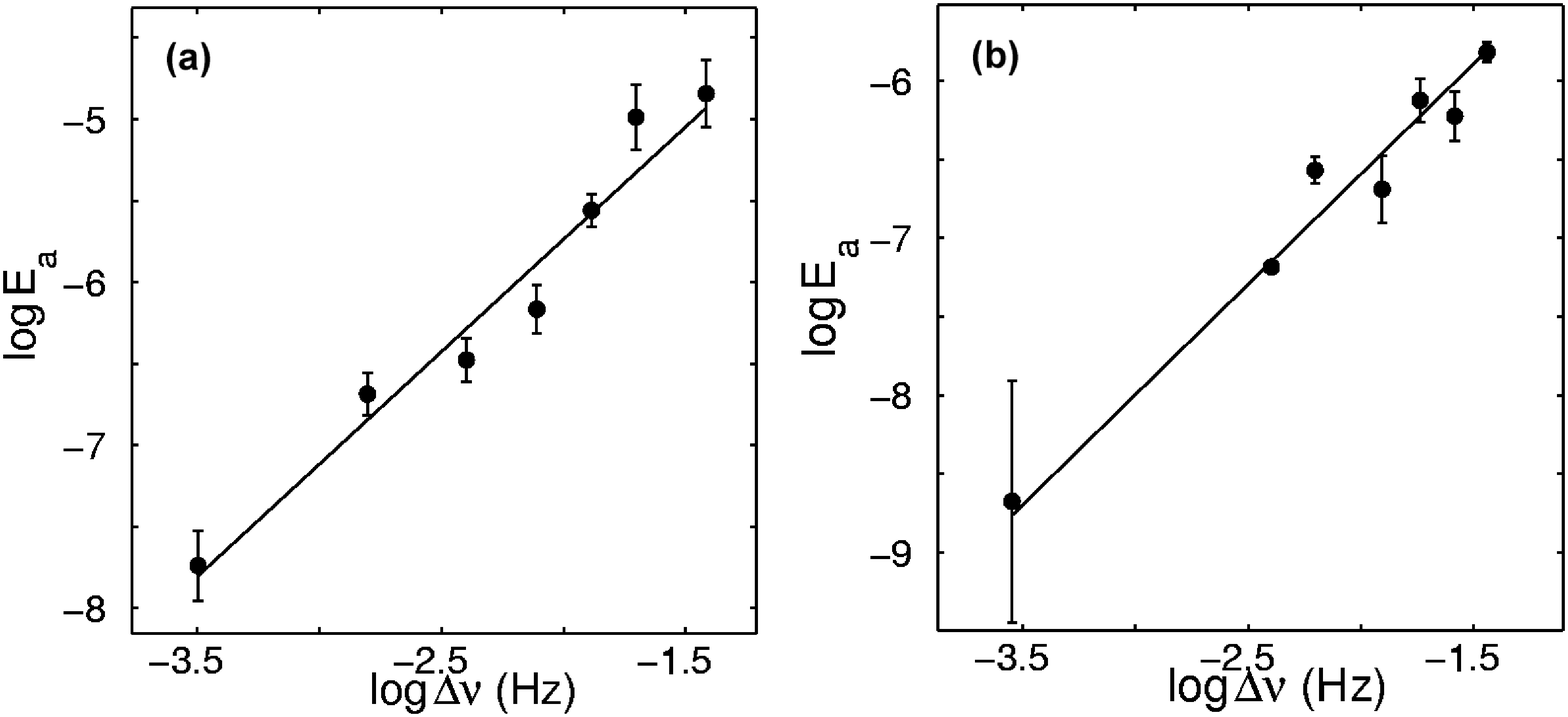} 
\caption{\label{fig:4} Dependence of the activation energy on detuning frequency for (a) sample A and (b) sample B.  The solid lines are power law fits, yielding critical exponents of $1.38 \pm 0.15$ and $1.4 \pm 0.15$ respectively.}
\end{figure}

The strong nonlinearity in our torsional oscillator derives mainly from the distance dependent electrostatic interaction between the top plate and the excitation electrode. The equation of motion of the oscillator is given by \cite{17}:
\begin{equation}\label{eq:1}
         \ddot{\theta} + 2 \gamma  \dot{\theta}  + \omega_{0}^{2} \theta  = \frac{\tau}{I}, 
\end{equation}
where $\theta$ is the angular rotation of the top plate, $\gamma$ is the damping coefficient, $\omega_0$ is the natural frequency of the oscillator and $\tau$ is the driving torque. The driving torque arises from the electrostatic interaction between the top plate and the driving electrode. For small rotation angle $\theta$, $\tau$ can be written as:
\begin{equation}\label{eq:2}
         \tau = bF, 
\end{equation}
where $b$ ($\sim125\; \mu m$) is half the distance from the axis of rotation to the edge of the top plate. The electrostatic force $F$ is evaluated with separation  $d-b\theta$:
\begin{equation}\label{eq:3}
F=\frac{\epsilon_{0}AV^2}{2(d-b\theta)^2}.
\end{equation}
As shown in Fig.\ \ref{fig:1}a, $d$ ($\sim2\; \mu m$) is the separation between the top plate and the electrode in the absence of external torque, $A$ is the area of the electrode and $\epsilon_0$ is the permittivity of free space. The excitation voltage $V$ consists of three components:
\begin{equation}\label{eq:4}
	V = V_{dc1}+V_{ac}sin(\omega t) +V_{noise}(t).
\end{equation}
The three terms on the right side of Eq.\ (\ref{fig:4}) represent the dc voltage, periodic ac voltage with angular frequency $\omega$ and random noise voltage respectively. $V_{dc1}$ is chosen to be much larger than $V_{ac}$ and $V_{noise}$ to linearize the dependence of $F$ on $V_{ac}$ and $V_{noise}$. The strong distance dependence of the electrostatic force leads to nonlinear contributions to the restoring torque. A Taylor expansion of the electrostatic force $F$ about $d$  in Eqs.\ (\ref{eq:1}) and (\ref{eq:2}) leads to:
\begin{equation}\label{eq:5}
         \ddot{\theta} + 2 \gamma  \dot{\theta}  + [\omega_{0}^{2} - \eta]\theta + \alpha \theta^2 + \beta \theta^3 - C = Esin(\omega t) +N(t), 
\end{equation}
where $\alpha=-\frac{3b^3\epsilon_0A}{2Id^4}V_{dc1}^2$, $\beta=-\frac{2b^4\epsilon_0A}{Id^5}V_{dc1}^2$, $\eta=\frac{b^2\epsilon_0A}{Id^3} V_{dc1}^2$, $C=\frac{b\epsilon_0A}{2Id^2} V_{dc1}^2$,  $E=\frac{b\epsilon_0AV_{dc1}}{2Id^2} V_{ac}$ is the effective amplitude of the periodic excitation and  $N(t)=\frac{b\epsilon_0AV_{dc1}}{2Id^2} V_{noise}(t)$ is the effective noise in the excitation. The linear term in the Taylor expansion of $F$ produces a shift in the natural frequency of oscillation, while the nonlinear effects arising from $\alpha$ and $\beta$ are characterized by a constant \cite{17} $\kappa=3\beta/8\omega_0 - 5\alpha^2/12{\omega_{0}}^2$. For our device geometry, $\beta/\alpha\approx330$ and $\beta/{\omega_0}^2 \approx10000$. As a result, the contribution of the quadratic nonlinearity (the $\alpha$ term) to $\kappa$  is negligible compared to the cubic nonlinearity (the $\beta$ term) and our device can be regarded as a Duffing oscillator in the absence of injected noise. 

First, we focus on the response of the oscillator with no injected noise in the excitation. We show in Fig.\ \ref{fig:1}b the frequency response of sample A at two different excitation amplitudes. Both responses have been normalized so their respective maximum amplitudes are one. The squares represent the response of the oscillator at the smaller excitation. The resonance peak is fitted well by the dotted line that corresponds to the response of a damped harmonic oscillator. As the periodic excitation is increased, the cubic term in Eq.\ (\ref{eq:5}) leads to nonlinear behavior in the oscillations. The resonance curve becomes asymmetric with the peak shifting to lower frequencies, consistent with a negative value of $\beta$. At a high enough excitation, hysteresis occurs in the frequency response, as shown by the circles in Fig.\ \ref{fig:1}b. Within a certain range of driving frequencies, there are two stable dynamic states with different oscillation amplitude and phase. Depending on the history of the oscillator, the system resides in either the high-amplitude state or low-amplitude state. In the absence of fluctuations, the oscillator remains in one of the stable states indefinitely. 

When sufficient noise is applied in the excitation, the oscillator is induced to escape from one dynamic state into the other. Since this driven, bistable system is far from thermal equilibrium and cannot be characterized by free energy, calculation of the escape rate is a non-trivial problem. Theoretical analysis \cite{14,16} suggests that the rate of escape $\Gamma$ at a particular driving frequency depends exponentially on the ratio of an activation energy $E_a$ to the noise intensity $I_N$:
\begin{equation}\label{eq:6}
       \Gamma \propto  e^{-E_a/I_N}. 
\end{equation}
Close to the bifurcation frequency, where the high-amplitude state disappears, the activation energy is expected to display system-independent scaling:
 \begin{equation}\label{eq:7}
         E_a\propto(\nu - \nu_{b})^{\rho}\propto(\Delta \nu)^{\rho}, 
\end{equation}
where the detuning frequency $\Delta \nu$ is the difference between the driving frequency
$\nu$ and the bifurcation frequency $\nu_{b}$. The activation energy is predicted \cite{14,16} to increase with frequency detuning with critical exponent $\rho = 3/2$. To our knowledge, this scaling relation has not yet been observed, even though it is expected to occur in a number of non-equilibrium systems. We describe below our comprehensive experimental investigation of activated switching from the high-amplitude to the low-amplitude state for two micromechanical oscillators with different resonant frequencies and damping coefficients. The critical exponents measured for both samples were in good agreement with theory.

\begin{figure}[ht]
\includegraphics[scale=.21,angle=0]{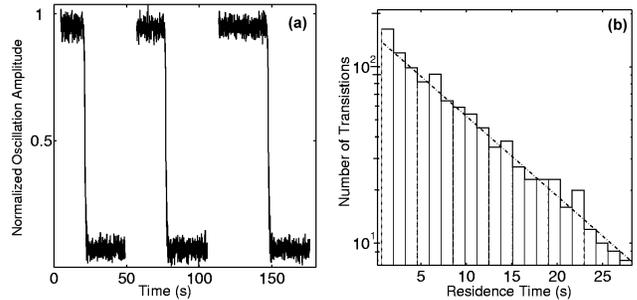} 
\caption{\label{fig:2} (a) In the presence of noise in the excitation, the oscillator switches from the high-amplitude state to the low-amplitude state at different time intervals. The system is reset to the high-amplitude state between switching events. The detuning frequency is $\Delta \nu =0.05\; Hz$.
(b) Histogram of the residence time in the upper state before switching occurs, at a detuning frequency of $0.05\; Hz$ for sample A. The dotted line is an exponential fit. 
}
\end{figure}

In our experiment, we induce transitions from the high-amplitude state to the low-amplitude state by injecting noise in the excitation with a bandwidth of $100\; Hz$  centered about the resonant frequency. The bandwidth of the noise is much larger than the width of the resonance peak. We chose a sufficiently large sinusoidal excitation so that the hysteresis loop exceeds twice the resonance peak width. Figure \ref{fig:2}a shows typical switching events at an excitation frequency of $3278.81\; Hz$ for sample A where the oscillator resides in the high-amplitude state for various durations before escaping to the low-amplitude state. Due to the random nature of the transitions, a large number of switching events must be recorded to determine the transition rate accurately. During the time interval between switching events in Fig.\ \ref{fig:2}a, the oscillator is reset to the high-amplitude state using the following procedure. First, the noise is turned off and the driving frequency is increased beyond the range of frequencies where bistability occurs ($>3280.8\; Hz$ as shown in Fig.\ \ref{fig:1}b). The driving frequency is then decreased slowly towards the target frequency so that the oscillator remains in the high-amplitude state. Once the target frequency is reached, the noise is turned back on and the time for the oscillator to escape from the high-amplitude state is recorded. This process is then repeated multiple times to accumulate the statistics for switching. Such a procedure is necessary because the energy barrier for transitions from the low-amplitude state back to the high-amplitude state is much larger than the barrier for transition in the opposite direction. Thus, noise induced transitions from the low-amplitude state to the high-amplitude state will fail to occur in the duration of the experiment and the oscillator must be reset to the high-amplitude state using the steps described above. Figure \ref{fig:2}b shows a histogram of the residence time in the high-amplitude state before a transition occurs. The exponential dependence on the residence time indicates that the transitions are random and follow Poisson statistics as expected.

\begin{figure}[ht]
\includegraphics[scale=.29,angle=270]{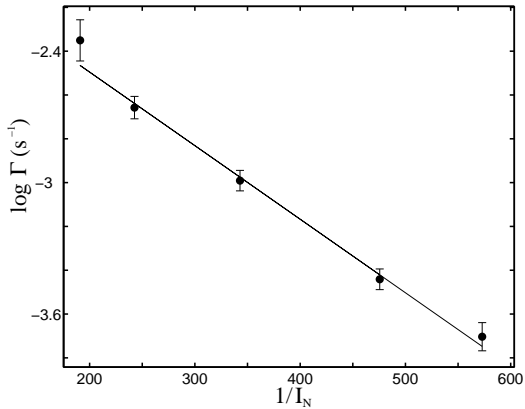} 
\caption{\label{fig:3} Logarithm of the transition rate from the high-amplitude state as a function of inverse noise intensity at a detuning frequency $\Delta \nu$ of $0.25\; Hz$ for sample A. The slope of the linear fit yields the activation energy.
}
\end{figure}

To determine the activation energy for a particular detuning frequency, we record a large number of transitions for multiple noise intensities ($I_N$). The average residence time at each noise intensity is extracted from the exponential fit to the corresponding histograms. Figure \ref{fig:3} plots the logarithm of the average transition rate as a function of inverse noise intensity. The transition rate varies exponentially with inverse noise intensity, demonstrating that escape from the high-amplitude state is activated in nature. According to Eq.\ (\ref{eq:6}), the slope in Fig.\ \ref{fig:3} yields the activation energy for escaping from the high-amplitude state at the particular detuning frequency.

We repeat the above procedure to determine the activation energy for other detuning frequencies $\Delta \nu$ ($\Delta \nu$ is the difference between the driving frequency and the bifurcation frequency at which the high-amplitude state disappears) and show the result in Fig.\ \ref{fig:4}a for sample A and Fig.\ \ref{fig:4}b for sample B. All the detuning frequencies chosen for sample A are smaller than its resonance peak width while the maximum detuning frequency for sample B is about $4$ times its resonance peak width. Fitting the activation energies with a power law dependence on the detuning frequency yields critical exponents of $1.38 \pm 0.15$   for sample A and $1.40 \pm 0.15$  for sample B respectively. Despite the different resonant frequencies and a factor of $4$ difference in damping, the critical exponents obtained for both samples are in good agreement with theoretical predictions \cite{13,14,16}. To our knowledge, this represents the first experimental verification of universal scaling for activated escape close to saddle-node bifurcations in systems far from thermal equilibrium.
 
We remark that recently Aldridge and Cleland \cite{11} measured noise-induced switching between dynamical states in a nanomechanical beam. They found a quadratic dependence of the activation barrier on the distance to the critical point where the two stable states of forced vibrations and the unstable periodic state all merge together. Such quadratic dependence arises \cite{14} when the parameters are changed along the line where the populations of the two stable states are equal to each other. This requires changing simultaneously both the amplitude and the frequency of the driving field. In contrast, in our experiment, we approach a bifurcation point where a stable large-amplitude state and the unstable state merge together, while the stable small-amplitude state is far away. We vary only one parameter, the detuning frequency, while maintaining the periodic driving amplitude constant. We found that the activation barrier for escape is reduced to zero with critical exponent of $3/2$.

In conclusion, we demonstrated the activated behavior of noise induced switching of a thermally nonequilibrium system, a nonlinear underdamped micromechanical torsional oscillators modulated by a strong resonant field. The measured critical exponent for the activation energy near the bifurcation point agrees well with the predicted value of $3/2$, verifying the system-independent scaling of the activation energy in the vicinity of the bifurcation point. Other systems that are far from equilibrium, including RF-driven josephson junctions \cite{12,20}, nanomagnets driven by polarized current \cite{21} and double barrier resonant tunneling structures \cite{22}, are expected to obey the same scaling relationship near the spinodal point.

We thank M. I. Dykman and D. Ryvkine for useful discussions.


\end{document}